\documentclass[journal]{IEEEtran}
\usepackage{amsmath,amsfonts, amssymb}
\usepackage{algorithmic}
\usepackage{algorithm}
\usepackage{array}
\usepackage[caption=false,font=normalsize,labelfont=sf,textfont=sf]{subfig}
\usepackage{textcomp}
\usepackage{stfloats}
\usepackage{url}
\usepackage{verbatim}
\usepackage{graphicx}
\usepackage{cite}
\usepackage{orcidlink}
\usepackage{upgreek}
\usepackage{hyperref}
\hypersetup{
	colorlinks=true,
	linkcolor=blue,
	filecolor=magenta,      
	urlcolor=cyan,
	citecolor=red, 
}

\graphicspath{{Figures/}}

\begin{document}
	
	\title{Investigating the performance of RPM JTWPAs by optimizing LC-resonator elements
	}
	\author{M. A. Gal\'i Labarias~\orcidlink{0000-0002-8934-0006}
		~\thanks{Correspondence should be sent to galilabarias.marc@aist.go.jp},
		T. Yamada,~\orcidlink{0000-0002-2323-5609},
		Y. Nakashima,~\orcidlink{0000-0001-9783-6784},
		Y. Urade,~\orcidlink{0000-0002-6580-5497},
		K. Inomata~\orcidlink{0000-0002-1151-4512}
				\thanks{The authors are with the Global Research and Development Center for Business by Quantum-AI technology (G-QuAT) at the National Institute of Advanced Industrial Science and Technology (AIST), Tsukuba, Ibaraki, Japan.}
		}
	
%
	
	\maketitle
	
	\begin{abstract}
		Resonant phase-matched Josephson traveling-wave parametric amplifiers (RPM JTWPAs) play a key role in quantum computing and quantum information applications due to their low-noise, broadband amplification, and quadrature squeezing capabilities.
		This research focuses on optimizing RPM JTWPAs through numerical optimization of parametrized resonator elements to maximize gain, bandwidth and quadrature squeezing. Our results show that optimized resonators can increase the maximum gain and squeezing by more than 5 dB in the ideal noiseless case. However, introducing the effects of loss through a lumped-element model reveals that gain saturates with increasing loss, while squeezing modes degrade rapidly, regardless of resonator optimization. 
		These results highlight the potential of resonator design to significantly improve amplifier performance, as well as the challenges posed by current fabrication technologies and inherent losses.		 
	\end{abstract}
	
	
\section{Introduction}

Fabricating Josephson traveling-wave parametric amplifiers (JTWPAs) presents several challenges due to circuit design and the fabrication complexity of their constitutive parts~\cite{Yurke1988, Yamamoto2008,  Castellanos2007, Mallet2011, Esposito2021, Kissling2023, Malnou2025, Gaydamachenko2025, Sandbo2025}. Therefore, mathematical models offer a unique platform to study these nonlinear devices, since they allow us to numerically, and sometimes analytically, solve the wave equations describing the dynamics of the system~\cite{Yurke1989, Chen1989, Cullen1960}. 
In recent years many theoretical approaches have been introduced, from studying the classical response and amplification of three- and four-wave mixing~\cite{Yaakobi2013, OBrien2014, Zorin2016, Peng2022_PRX, Dixon2020, Gaydamachenko2022, Malnou2024a}; to quantized models which allow to study the device squeezing~\cite{Grimsmo2017, Qiu2023}.

In the present work we numerically optimize the resonator parameters of a resonant phase-matched Josephson traveling-wave parametric amplifier (RPM JTWPA). We first introduce a quantized model for the wave modes which allow us to study the classical (gain) and quantum (squeezing) response of the RPM JTWPA~\cite{OBrien2014, Grimsmo2017, Gali2025}. Then, we define an upper bound on the effects of loss by adding a beam splitter at the end of the device through a phenomenological model~\cite{ Grimsmo2017, Houde2019}.
Finally, we parametrize the resonator elements in order to perform numerical analysis on the gain and squeezing responses.

\section{Method}
In this section we give a summarized overview of the main steps followed to derive the mathematical model of the {RPM JTWPA} investigated in this work.
Here, we closely follow models previously introduced in the literature and refer the reader to them for details on the derivation~\cite{Santos1995, OBrien2014, Grimsmo2017, Gali2025}.

\begin{figure}
	\centering
	\includegraphics[scale=0.35]{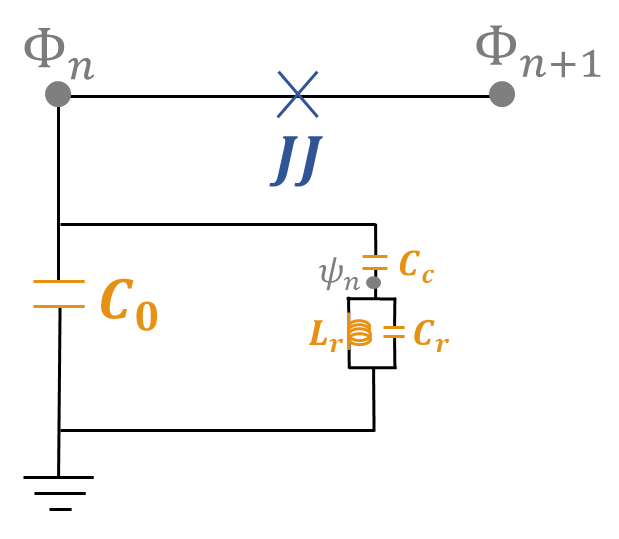}
	\caption{Unit-cell diagram of an RPM JTWPA. Gray dots labeled by $\Phi_{n}$, $\Phi_{n+1}$ and $\Psi_n$ are node fluxes~\cite{Devoret1995}. The blue cross indicates the Josephson junction,  $C_{0}$, $C_{c}$ and $C_{r}$ are the capacitances and $L_r$ is the resonator inductance. Gold color depicts the resonator's parameters which will be investigated here.}
	\label{fig:RPM_JTWPA_UC}
\end{figure}

The system under study is described by the unit cell depicted in Fig.~\ref{fig:RPM_JTWPA_UC}, in other words a JTWPA will be composed of $N_{\text{JJ}}$ copies of this unit cell, where $N_{\text{JJ}}$ is the total number of JJs in the device. Under the continuous approximation, its Lagrangian density can be expressed in normalized units as follows~\cite{Gali2025}, 
\begin{align}\label{eq:Lagragian_den_JTWPA}
	\mathcal{L}(\phi, \psi, \dot{\phi}, \dot{\psi}) &= \frac{E_{J}}{2} \Big\{ c_0 \dot{ \phi}^2 +
	 \left( \partial_x \dot{\phi} \right)^2 
	+ c_c \left(  \dot{\psi} - \dot{\phi} \right)^2 \nonumber \\
	&+ c_r  \dot{\psi}^2 + 2 \cos(\partial_x \phi) - \frac{1}{l_r }\psi^2 \Big\}  
	\, , 
\end{align}
where lowercase letters indicate normalized variables, such that: $c_0 := C_0 / C_J$ with
$C_J$ the JJ capacitance; $l_r := L_r / L_J$ where $L_J$ and $L_r$ are the JJ and resonator inductances, respectively; $\phi := \Phi / \Phi_0$ and $\psi := \Psi / \Phi_0$ are normalized node fluxes~\cite{Devoret1995} with $\Phi_0$ the flux
quantum. The Josephson energy is $E_J = I_c \Phi_0/(2\pi)$ with
$I_c=\Phi_0/(2\pi L_J)$ the JJ critical current (assuming identical
JJs across the device). The symbol $\partial_x$ is the partial derivative in the spatial dimension normalized by the unit-cell length. The dot indicates the time derivative over the normalized time $\tau :=\omega_J t$ where ${\omega_J = (L_J C_J)^{-1/2}}$ is the Josephson plasma frequency.

In the frequency domain, Eq.~\eqref{eq:Lagragian_den_JTWPA} can be solved analytically under the strong-pump approximation and expanding $\phi$ in three modes: pump, signal and idler.
The magnitude of the signal mode is found to be
\begin{equation}
	\left| \hat{a}(x, \omega) \right|= 
	 \left| u(x, \omega) \hat{a}(0, \omega) + i v(x, \omega) \hat{a}^{\dagger}(0, 2\omega_p - \omega) \right| \, ,
	\label{eq:JTWPA_sol1}
\end{equation}
with $\omega_p$ the pump frequency and $i=\sqrt{-1}$ the imaginary unit.

The complex functions $u(x, \omega)$, $v(x, \omega)$ and $g(\omega)$ depend on the pump power and operating frequencies and are defined as follows,
\begin{align}
	u(x, \omega) &= \cosh \left[ g(\omega) \, x \right] - i \frac{\Delta k(\omega)}{2g(\omega)}\sinh \left[ g(\omega) \, x \right] \, ,\\
	v(x, \omega) &= \frac{	\beta^2 }{g(\omega)} \sqrt{k(\omega) k(2\omega_p - \omega)} \sinh \left[ g(\omega) \, x \right] \, ,\\
	g (\omega) &= \sqrt{ \beta^4 k(\omega) k(2\omega_p - \omega) - \left( \frac{\Delta k (\omega)}{2 } \right)^2} \, , 
\end{align}
where 
 ${\Delta k =  \left( 1 + 2|\beta|^2 \right) \Delta k_L  - 2|\beta|^2 k (\omega_p) }$ is the total phase difference between pump, signal and idler, with ${\Delta k_L = 2k (\omega_p) - k (\omega) - k (2\omega_p - \omega)}$ the linear phase difference and $\beta = I_p / (4 I_c)$ the normalized pump current with $I_p$ the pump current.

With this formulation the signal gain in decibels has the following closed expression
\begin{align}\label{eq:Gain}
	G(x, \omega) := 10 \log_{10} \left| u(x, \omega) \right|^2 \, ,
\end{align}
where we have assumed that the idler initial amplitude, $|\hat{a}(0, 2\omega_p - \omega)|$, is negligible compared to that of the signal.

\subsection{Beam-Splitter Loss Model}
A simple method to incorporate the effects of loss is to add a beam-splitter~\cite{Houde2019, Grimsmo2017} at the output of the JTWPA with a transmittance $\eta$.
Here $\eta$ is a constant, thus this model does not account for loss asymmetry, which it has been shown~\cite{Houde2019} that destroys squeezing when the average loss is large enough. Thus our current model produces an upper bound on squeezing, since including loss along the transmission line and allowing for loss asymmetry will always produce less gain and squeezing than our current model.
 
With these consideration, using Eq.~\eqref{eq:JTWPA_sol1} with the beam splitter at the end of the device, the final output modes will be defined as,
\begin{equation}
	\hat{a}_{\xi}(N_{\text{JJ}}, \omega) = \sqrt{\eta} \hat{a}(N_{\text{JJ}}, \omega) + \sqrt{1 - \eta} \hat{\xi}(\omega) \, ,
\end{equation}
where $\hat{\xi}(\omega)$ represents the noise modes created due to reflection at the beam-splitter, with reflectance coefficient $1-\eta$. Note that to be well defined the noise operators $\hat{\xi}$ must satisfy the usual commutation relationships of the creation/annihilation operators.

We now define the canonical quadrature at the end of the JTWPA, i.e. $x=N_{\text{JJ}}$, as follows:
\begin{align}
	\hat{X}_{\theta} (\omega) &= e^{i \theta/2} \hat{a}^{\dagger}_{\xi} (N_{\text{JJ}}, \omega) + e^{- i \theta/2} \hat{a}_{\xi} (N_{\text{JJ}}, \omega)\, ,
\end{align}
and its squeezing spectrum as
\begin{align*}
	S_{X_{\theta}}(\omega)&  = \int_{-\infty}^{\infty} d\omega' \langle \hat{X}_{\theta}^{\dagger}(\omega) \hat{X}_{\theta}(\omega')\rangle \\
	&= 2\eta |u(N_{\text{JJ}}, \omega ) v(N_{\text{JJ}}, \omega)| \sin \theta  +  2\eta |v(N_{\text{JJ}}, \omega)|^2 + 1  \, .
\end{align*}

At $\theta = -\pi/2$ the quadrature-squeezing is maximized. Thus, we define the squeezing as,
\begin{align}\label{eq:S}
	S_{X}(\omega):= -2\eta | u(N_{\text{JJ}}, \omega) v(N_{\text{JJ}}, \omega )| +  2\eta |v(N_{\text{JJ}}, \omega)|^2 + 1  \, .
\end{align}

\subsection{Resonator Elements Parametrization}

The effective impedance of the resonator described in Fig.~\ref{fig:RPM_JTWPA_UC} is
\begin{subequations}
	\begin{align}\label{eq:RPM-Z_eff}
		Z_{\text{eff}}^{-1}  &= i \omega C_{0} + i \omega  C_{c}\frac{ 1 - \omega^2 C_{r} L_{r}}{ 1 -  \omega^2 (C_{c} + C_{r} ) L_{r} }  \, .
	\end{align}
\end{subequations}

Thus its resonance frequency occurs at,
\begin{align}
	\omega_{r} &= \left[ (C_{c} + C_{r} ) L_{r} \right]^{-1/2} 
	\, . \label{eq:RPM_freq}
\end{align}
Away from resonance and using that $C_{c} \ll C_{r}$ then ${Z_{\text{eff}}^{-1} \approx i \omega C_{\text{eff}} }$ where we have defined 
\begin{equation}\label{eq:C_eff}
C_{\text{eff}} :=  C_0 + C_{c} \, .	
\end{equation}

As the RPM condition depends on the resonant frequency, Eq.~\eqref{eq:RPM_freq}, there are an arbitrary combination of $\{L_{r}, C_{r}, C_c\}$ which gives the same resonant frequency.
Similarly, by fixing a unit-cell impedance $Z_0=50 \, \Omega$, the relationship between $C_0$ and $C_c$ is determined by $Z_0^2 = L_J / C_{\text{eff}} $ and the constrain that capacitances cannot be negative, i.e. $C_c, \, C_r \ge 0$.
Therefore, for a fixed resonator frequency and unit-cell impedance, the system has two degrees of freedom.
This naturally rises the following question: \emph{Do these two degrees of freedom affect the RPM JTWPA gain and squeezing spectra?}

\section{Results and Discussion}

To analyze how the resonator parameters affect the device response we will use $C_c$ and $C_r$ as our free parameters, with $L_r$ and $C_0$ being determined by Eqs.~\eqref{eq:RPM_freq} and \eqref{eq:C_eff} when resonator frequency and unit-cell impedance are fixed.
For the following numerical simulations we will use these parameter values: number of unit cells $N_{\text{JJ}}=2000$, unit-cell characteristic impedance $Z_0 = 50\, \Omega$,  JJ critical current ${I_c =2.75 \, \upmu}$A, $C_J=39.5\,$fF ($\omega_J/(2\pi)= 73.17\,$GHz), pump current ${I_p=1.37 \upmu}$A and resonator frequency $f_r=6.06 \,$GHz.

It is important to note that in this paper we are considering a resonator every unit cell, however fabricated devices usually include a resonator periodically every few unit cells~\cite{Macklin2015, White2015, Qiu2023}, in order to facilitate fabrication and to reduce dielectric loss due to the resonator capacitances. Having resonators added periodically, instead of at every unit cell, decreases phase matching but also loss. This trade-off creates an optimization problem: determining the number of resonators that provides the best performance, and it depends on the materials and technology used to fabricated those. This is a different optimization problem and would require an extension of a distributed-loss model~\cite{Caves1987, Ranadive2022, Qiu2023}; this is outside the scope of the present work, which focuses on optimizing the capacitance values of the resonator.

\subsection{Ideal case: zero loss}

\begin{figure}
	\centering
	\includegraphics[width=0.5\textwidth]{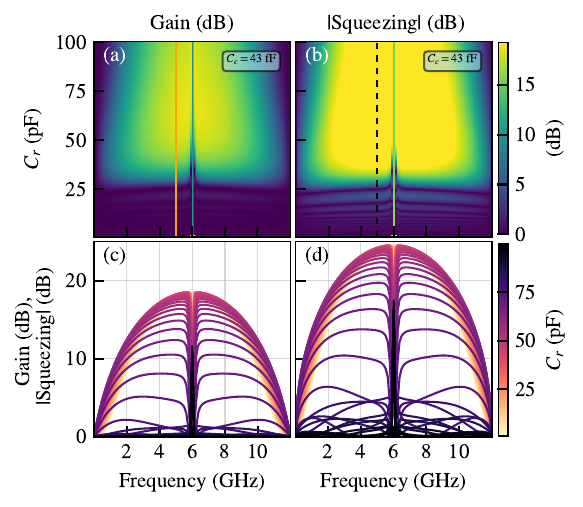}
	\caption{Gain [(a) and (c)] and squeezing absolute value [(b) and (d)] depending on the signal frequency and resonator capacitance, $C_r$, at a fixed ${C_c=43\,}$fF and $f_p = 6 \,$GHz.}
	\label{fig:GainOpt}
\end{figure}

\begin{figure}
	\centering
	\includegraphics[width=0.5\textwidth]{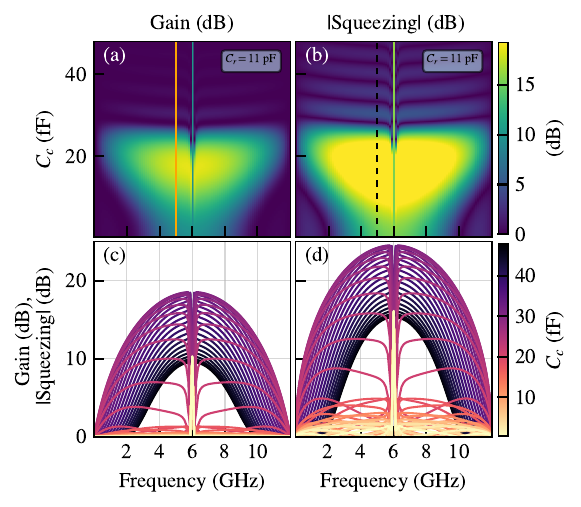}
	\caption{Gain [(a) and (c)] and squeezing absolute value [(b) and (d)] depending on the signal frequency and resonator coupling capacitance, $C_c$, at a fixed ${C_r=11\,}$pF  and $f_p = 6 \,$GHz.}
	\label{fig:GainOptB}
\end{figure}

\begin{figure}
	\centering
	\includegraphics[width=0.45\textwidth]{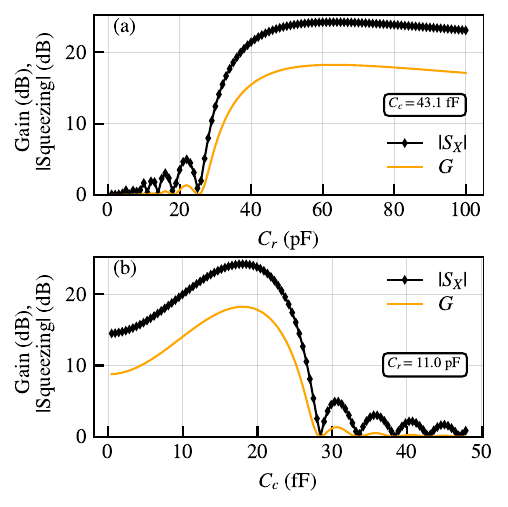}
	\caption{Gain (yellow line) and absolute value of the squeezing (black line with diamonds) at 5 GHz  and $f_p = 6 \,$GHz depending on: (a) the resonator capacitance for a fixed coupling capacitance $C_c=43.1\,$fF, and (b) the coupling capacitance for a fixed $C_r=11 \,$pF. These lines correspond to the same colored lines in Fig.~\ref{fig:GainOpt}.}
	\label{fig:cross_sec}
\end{figure}

\begin{figure}
	\centering
	\includegraphics[width=0.45\textwidth]{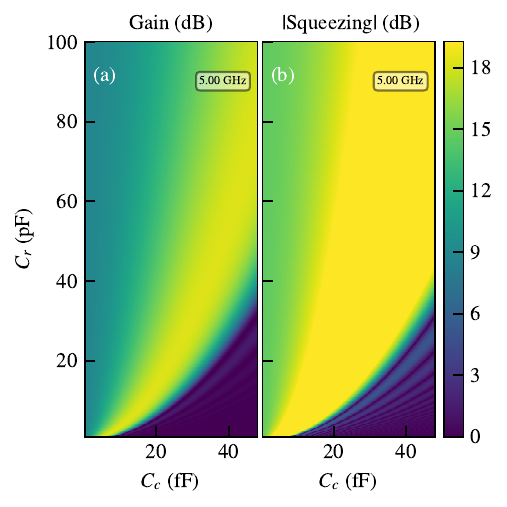}
	\caption{(a) Gain and (b) squeezing absolute value at ${f=5\,}$GHz depending on $C_c$ and $C_r$. Pump frequency has been optimized to achieve maximum gain at each point.}
	\label{fig:CrCcheatmap}
\end{figure}

\begin{figure}
	\centering
	\includegraphics[width=0.45\textwidth]{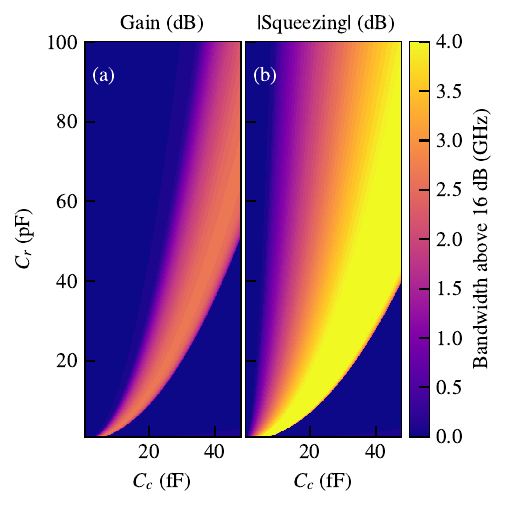}
	\caption{Bandwidth above 16 dB depending on $C_c$ and $C_r$ for (a) $G$ and (b) $|S_X|$. Pump frequency has been optimized to achieve maximum gain at each point.}
	\label{fig:CrCcheatmap_BW}
\end{figure}

\begin{figure}
	\centering
	\includegraphics[width=0.45\textwidth]{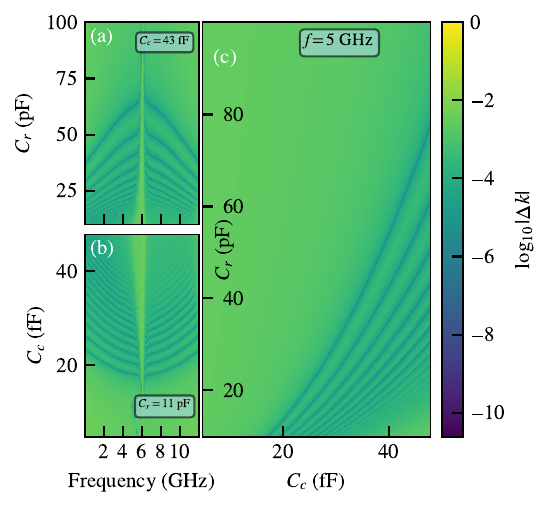}
	\caption{Phase difference depending on signal frequency at fixed $C_c$ (a) and fixed $C_r$ (b). (c) depicts the phase difference at a fixed frequency 5 GHz depending on $C_c$ and $C_r$. Pump frequency has been optimized to achieve maximum gain at each point.}
	\label{fig:CrCcheatmap_Dk}
\end{figure}

When designing the elements of a resonator, one might have certain limitations due to, for example, fabrication technologies, materials used or circuit design. These can restrain our options, so a suboptimal value is used. In Figs.~\ref{fig:GainOpt}-\ref{fig:cross_sec} we show that for a not optimal choice of $C_c$ ($C_r$) there is a range of $C_r$ ($C_c$) that can still optimize the response. In Figs.~\ref{fig:CrCcheatmap}-\ref{fig:CrCcheatmap_Dk} we optimize the resonator elements by spanning over a range of $(C_r, C_c)$ combinations each pumped at an optimized pump frequency.

Figure~\ref{fig:GainOpt} shows the gain, (a) and (c), and squeezing spectra, (b) and (d), depending on $C_r$ for a fixed ${C_c=43\,}$fF and $f_p = 6 \,$GHz.
In Figs.~\ref{fig:GainOpt}(a) and (b) show small performance at small $C_r<25\,$pF, while above that value $G$ and $S_X$ quickly increase. However, after achieving maximum performance, $\sim 50\,$pF, the device bandwidth starts to decrease.
The cross sections of Figs.~\ref{fig:GainOpt}(a) and (b) are shown in Figs.~\ref{fig:GainOpt}(c) and (d) respectively, for easier visualization.

In Figs.~\ref{fig:GainOpt}(c) and (d) $C_r=11\,$pF and $f_p = 6 \,$GHz are fixed, while $C_c$ is being changed. In this case, the device has poor performance at $C_c \gtrsim 25\,$fF, while performance increases at smaller $C_c$ values. Noticeably, the bandwidth degrades quickly after reaching an optimal point at $C_c \approx20\,$fF.

Alternatively, Fig.~\ref{fig:GainOptB} shows the gain and squeezing spectra depending on $C_c$ for a fixed $C_r=11\,$pF and $f_p = 6 \,$GHz; Figs.~\ref{fig:GainOptB}(c) and (d) show their respective cross sections.
In this case, the device has poor performance at $C_c \gtrsim 25\,$fF, while performance increases at smaller $C_c$ values. Noticeably, the bandwidth degrades quickly after reaching an optimal point at $C_c \approx20\,$fF.

Thus, while both $C_r$ and $C_c$ affect maximum gain and bandwidth, $C_c$ plays a stronger role in the device bandwidth, agreeing with previous studies~\cite{Navarro2021}.

Figure~\ref{fig:cross_sec} shows the cross sections marked by the solid yellow and dashed black lines in Fig.~\ref{fig:GainOpt}. It can be seen that both gain and squeezing have optimal point with $C_c$ being the most relevant parameter, as seen by the quick decrease of performance for $C_c$ away from the optimal point.

When resonator parameters are optimized, maximum performance is achieved at pump frequencies close to the resonance frequency of the device, Figs.~\ref{fig:GainOpt}-\ref{fig:cross_sec}.
However, we have shown that when operating the device at $f_p \sim f_r$ for large $C_c$ and small $C_r$, e.g. Fig.~\ref{fig:cross_sec}, the gain becomes purely imaginary, i.e. $\beta^4 k(\omega) k(2\omega_p - \omega) < \Delta k^2 (\omega) / 4$. In this regime the gain becomes sinusoidal and follows,
\begin{align}
	|u(x, \omega)|^2 &= 1 + \left(\epsilon - 1\right)^{-1} \sin^2 (|g(\omega)| x) \, ,
\end{align}
with $\epsilon:= \Delta k^2 \left( 4 \beta^4 k(\omega) k(2\omega_p - \omega) \right)^{-1}$, similarly the squeezing becomes a combination of sine and cosine terms.
This regime can be avoided by tuning the pump frequency away from resonance. At these pump frequencies the device performance is reduced compared to $f_p \sim f_r$, but allows us to continue operating the device with substantial gain and squeezing at all $(C_r, C_c)$ combinations.

To visualize the full parameter space, in Fig.~\ref{fig:CrCcheatmap} we plot $G$ and $S_X$ depending on both capacitances by fixing the signal frequency at 5 GHz and tuning $f_p$ GHz at each point to maximize gain. For these simulations we have discretized the pump frequency optimization range in gigahertz as ${f_{p,n} = 5.5 +  0.017n}$ with $n=0,\dots, 30$.

At small $C_r$ and $C_c$ values gain and squeezing are minimum.
The $(C_c, C_r)$-parameter space reveals multiple underlying trends which offer best performance. These trends are emphasized in Fig.~\ref{fig:CrCcheatmap_BW}, where the bandwidth above 16 dB is shown, and clearly depicted in Fig.~\ref{fig:CrCcheatmap_Dk} where these trends show best phase matching.

The interplay between $C_c$ and $C_r$ shown in Fig.~\ref{fig:GainOpt} - Fig.~\ref{fig:CrCcheatmap_BW}, can be better understood looking at Fig.~\ref{fig:CrCcheatmap_Dk}. 
Tuning $f_p$ to maximize gain, we plot the phase difference as a function of frequency [Fig.~\ref{fig:CrCcheatmap_Dk}(a) and (b)]; and as a function of $C_c$ and $C_r$ at a fixed frequency of 5 GHz, Fig.~\ref{fig:CrCcheatmap_Dk}(c).
Figures~\ref{fig:CrCcheatmap_Dk}(a)-(b) depicts a family of traces for which $\Delta k$ is minimized. As expected there is not a single pair of $C_c$ and $C_r$ which minimizes $\Delta k$ at all frequencies, however one can choose the pair that minimizes the phase-difference at the desired operational point.
Figure~\ref{fig:CrCcheatmap_Dk}(c) shows that at 5 GHz there is a wide range of values which minimized $\Delta k$, and follows the same trends shown in Fig.~\ref{fig:CrCcheatmap} and Fig.~\ref{fig:CrCcheatmap_BW}.
The traces showing minimum $\Delta k$ show different pump frequency regimes, with optimal $f_p$ being close to $f_r$ for small $C_c$ and high $C_r$ (top-left corner in Fig.~\ref{fig:CrCcheatmap_Dk}) and gradually decreasing for larger $C_c$ and smaller $C_r$ values.
Note that how many optimal $\Delta k$ lines appear depends on  how much one can tune the pump frequency.

\subsection{Upper bound on the effect of loss}

Modeling loss by adding a beam-splitter at the end of the JTWPA allows us to obtain an upper bound of the effect of loss in the gain and squeezing spectra, in other words, a best case scenario when loss appears.

\begin{figure}
	\centering
	\includegraphics[width=0.45\textwidth]{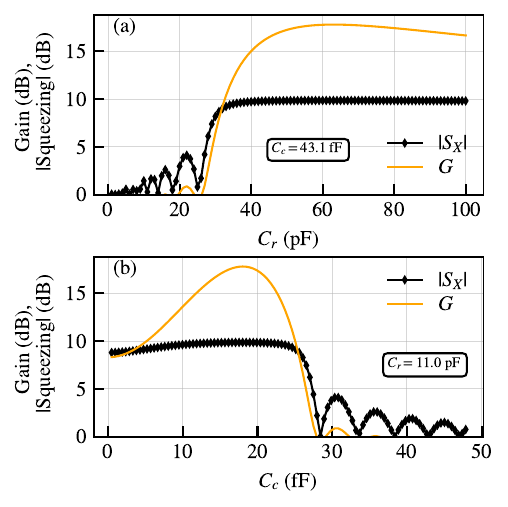}
	\caption{Gain (yellow line) and absolute value of the squeezing (black line with diamonds) at 5 GHz and with $\eta=0.9$ depending on (a) the resonator capacitance for a fixed coupling capacitance $C_c=43.1\,$fF; and (b) the coupling capacitance for a fixed $C_r=11 \,$pF.}
	\label{fig:eta_cross}
\end{figure}

Figure~\ref{fig:eta_cross} shows squeezing deterioration due to an estimated loss of 10\% ($\eta=0.9$), showing a saturation at 10 dB, while the gain reduction is small in comparison.
Even though parameter optimization is still relevant for the gain performance (yellow lines in Fig.~\ref{fig:eta_cross}); the maximum squeezing plateaus for large $C_r$, Fig.~\ref{fig:eta_cross}(a), and for small $C_c$, Fig.~\ref{fig:eta_cross}(b). 
This indicates that while gain can be further increased by design and fabrication optimization, squeezing spectra is mainly constrained by loss. Thus, to retain or improve squeezing performance one must prioritize device loss minimization.
Experimental squeezing measurements using nonlinear devices also showed that low loss is crucial to achieve entangled states and squeezing. Squeezing in a device with $\sim 0.6\,$dB loss has been demonstrated~\cite{Perelshtein2022}, this agrees with our results that squeezing would be degraded very quickly for $\eta<0.9$; for reference, a 0.6 dB loss in our framework corresponds to $\eta \approx 0.87$.

\begin{figure}
	\centering
	\includegraphics[scale=0.9]{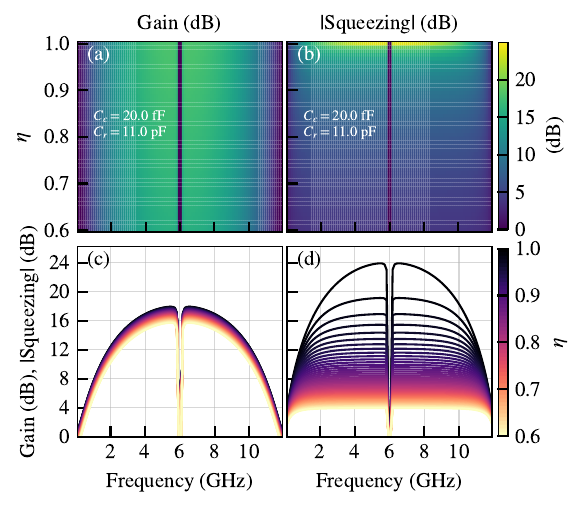}
	\caption{Loss effect in the gain (a) and squeezing (b) spectra. Their corresponding cross-sections are plotted in (c) and (d), respectively. Here the resonator values are fixed with $C_c=20 \,$fF and $C_r=11\,$pF, while $\eta$ is being swept.}
	\label{fig:eta}
\end{figure}

Fixing $C_c$ and $C_r$ close to optimal values we can investigate the response depending on $\eta$ and operating frequency.
Figure~\ref{fig:eta} shows the gain and squeezing for a device with ${C_c=20\,}$fF and ${C_r=11\,}$pF, which also has a resonance frequency ${f_r=6.06\,}$GHz.
Across the whole frequency spectrum, the overall response decreases monotonically with loss. While the gain reduction is relatively small, up to -2.2 dB for $\eta=0.6$, squeezing quickly decreases with increasing loss.

Note that this model only depicts an upper bound of the device performance under loss~\cite{Qiu2023}, this means that for a real device, where loss occurs along the transmission line, both gain and squeezing will be degraded more strongly than what is shown here~\cite{Esposito2022, Perelshtein2022, Qiu2023}.
If one is interested in studying all sources of loss in the measuring setup, one should consider the loss contribution from other electronics~\cite{Macklin2015, Malnou2024, Gaydamachenko2025}; for that a distributed-loss model would provide a better platform~\cite{Caves1987, Qiu2023}.

\section{Conclusion}

Our results showed that optimized resonators present peak and bandwidth improvement in gain and squeezing responses. 
While several resonator parameters can achieve similar maximum peak performance, bandwidth optimization requires a more careful selection of $(C_c, C_r)$ pairs (Fig.~\ref{fig:CrCcheatmap_BW}). We also showed that for most $C_c$ ($C_r$) values there is a $C_r$ ($C_c$) which maximizes performance; however, devices with high $C_c$ or small $C_r$ values require pump-frequency tuning at different operating frequencies to achieve maximum gain, while designs with smaller $C_c$ or larger $C_r$ values can be operated at a fixed and optimized pump frequency.

By adding a beam splitter at the end of the device, we can infer on the effects of loss on the performance of our device. The results presented here showed a slight decrease of gain, {-2.2 dB}; however, squeezing is rapidly reduced by loss. Interestingly, when loss is considered, the squeezing response plateaus showing similar responses for the different resonator element values analyzed. This trend emphasizes that squeezing is most sensitive to loss, and that resonator optimization is less relevant for squeezing performance than loss reduction, while for the gain response resonator optimization seems to be the driving process.
It is important to remind that the model presented here neglects higher idler and pump harmonics, thus even though we do not expect a variation in the optimal $(C_c, \, C_r)$ pairs or qualitative results; inclusion of higher harmonics would decrease the overall gain and squeezing, specially at high-pump power.

The results presented in this work indicate that resonator optimization can substantially improve the performance of a device. However, to optimize a real device one also needs to consider: (a) the number of resonators added per number of unit cells; (b) excitation of higher idler and pump harmonics; and (c) dielectric loss occurring along the transmission line. A model that includes these effects and balances number of resonators with their intrinsic loss is under development and will be the base of future publications.
	
\section*{Acknowledgments}
We thank Toyofumi Ishikawa for insightful and engaging discussions.
This work is based on results obtained from the project JPNP16007, commissioned by the New Energy and Industrial Technology Development Organization (NEDO), Japan.

%
%
%
%

	
	\bibliographystyle{IEEEtran}
	\bibliography{references_EUCAS2025}
	%

%
%
%
%
%
%
%
%
%
%
	
\end{document}